\documentstyle[multicol,aps,prl,twocolumn,epsf]{revtex}

\newcommand{\bm}[1]{\mbox{\boldmath $#1$}}
\begin{document}
\draft
\twocolumn[\hsize\textwidth\columnwidth\hsize\csname@twocolumnfalse\endcsname

\title{Statistical geometry in scalar turbulence}
\author{A. Celani and M. Vergassola}
\address{CNRS, Observatoire de la C\^ote d'Azur, B.P. 4229, 
06304 Nice Cedex 4, France}
\maketitle
\begin{abstract}

A general link between geometry and intermittency in passive scalar
turbulence is established. Intermittency is qualitatively traced back
to events where tracer particles stay for anomalousy long times in
degenerate geometries characterized by strong clustering. The
quantitative counterpart is the existence of special functions of
particle configurations which are statistically invariant under the
flow. These are the statistical integrals of motion controlling the
scalar statistics at small scales and responsible for the breaking of
scale invariance associated to intermittency.

\end{abstract}
]

Scalar fields transported by turbulent flow occur in many physical
situations, ranging from the dynamics of the atmosphere and the ocean
to chemical engineering (see, e.g., Ref.~\cite{SS00}). Specific
examples are provided by pollutant density, temperature or humidity
fields and the concentration of chemical or biological species. The
advection-diffusion equation governing the transport of the scalar
field $\theta$ is\,:
\begin{equation}
\label{passive}
\partial_t\theta({\bm r},t)+\left({\bm v}\cdot{\bm
\nabla}\right)\theta({\bm r},t)=\kappa\Delta\theta({\bm r},t),
\end{equation}
where ${\bm v}({\bm r},t)$ is the incompressible advecting flow and
$\kappa$ is the molecular diffusivity.  Two broad cases are
distinguished\,: active scalars, where ${\bm v}$ depends on $\theta$,
e.g. by an explicit relation ${\bm v}={\bm v}\left(\theta\right)$, and
passive scalars, where the statistics of ${\bm v}$ is independent of
$\theta$.  Here, we shall be concerned with the latter, although we
conjecture that the physical mechanisms presented in the following are
quite general and relevant also for the active cases. The
Fokker-Planck equation (\ref{passive}) is associated to the Lagrangian
dynamics of tracer particles whose position ${\bm \rho}(t)$ obeys
$d{\bm \rho}(t)={\bm v}({\bm \rho}(t),t)\,dt+\sqrt{2\kappa}\,\,d{\bm
\beta}(t)$, where ${\bm \beta}(t)$ is the isotropic Brownian motion
\cite{Risk}. The equation (\ref{passive}) governs the evolution of the
probability density of particles at position ${\bm r}$ and time $t$.

Scalar turbulence is typically generated by maintaining a mean scalar
gradient $\langle\theta\rangle={\bm g}\cdot{\bm r}$, e.g. by
heating/cooling devices in temperature field experiments.  The
notation $\langle\bullet\rangle$ denotes the average with respect to
the velocity statistics, which is in principle arbitrary.  We shall be
interested in flows with correlations having a nontrivial power law
behavior in the inertial range of scales $r\ll L$, where $L$ is the
velocity correlation length. Examples are provided by the two and
three dimensional Navier-Stokes turbulent flow (see, e.g.,
Ref.~\cite{UF95}).  A very robust feature of scalar turbulence is its
strong intermittency\,: rare strong events (such as the sharp cliffs
observed in Fig.~1) dominate the scalar statistical properties. More
quantitatively, intermittency reflects in the anomalous scaling of the
correlations. In the inertial range, the scalar structure functions
$S_{n}({\bm r})=\langle \left(\theta({\bm r},t)-\theta({\bm
0},t)\right)^n\rangle$ take the form
\begin{equation}
\label{strucfun}
S_{n}({\bm r})\propto r^{\zeta_n^{dim}}\,\,\left({L\over r}\right)
^{\zeta_n^{dim}-\zeta_n},
\end{equation}
with a nonvanishing value of $\zeta_n^{dim}-\zeta_n$. Here,
$\zeta_n^{dim}$ is the value predicted by simple mean field
dimensional arguments, e.g. of the Kolmogorov 1941 type.  The positive
value of the anomalous correction $\zeta_n^{dim}-\zeta_n$ reflects the
breaking of scale-invariance\,: the scale $L$ explicitly appears in
the inertial range expressions of scalar correlations, even though
$r\ll L$. The phenomenon of intermittency is quite generic for scalar
turbulence and independent of the specific choice of ${\bm v}$,
including for Gaussian velocity fields (see, e.g., Ref.~\cite{SS00}).
Past research on intermittency has mostly concentrated on
phenomenological models making vague contacts with the dynamics and
based uniquely on scaling exponents and structure functions. Angular
dependencies and geometrical informations about the correlations were
discarded. It is shown here that the whole structure of multi-point
correlations is in fact needed to achieve a real dynamical
understanding based on the geometry of the figures identified by
tracer particles.  Deviations of exponents from na\"{\i}ves
dimensional values turn out to be just by-products of the figure
geometry nontrivial evolution under Lagrangian dynamics.  While the
figure size grows according to simple dimensional arguments, the
evolution of their shape is a more delicate issue.  There are in
particular some special functions of the particle positions whose
average with respect to the Lagrangian dynamics remains constant in
time. That is due to a delicate compensation between the growth due to
the figure size and the depletion associated to the figure
shape. Those statistically conserved functions are dominating the
behavior in the inertial range and controlling the scalar field
intermittency.

Specifically, the velocity field ${\bm v}$ considered here is a
two-dimensional turbulent flow generated by an inverse energy cascade
process \cite{RHK67}. This is a situation of interest both for
experiments \cite{PT98,R98} and in the atmosphere \cite{Gage,Lind}.
The flow realizes the type of turbulence theoretically postulated by
A.N.~Kolmogorov in 1941\,: it is isotropic, it has a constant energy
flux (but upscale) and it is scale-invariant with scaling exponent
$1/3$ \cite{PT98,BCV00}. A property of interest to us is that the
velocity is not intermittent.  All nontrivial scaling properties of
the scalar field presented in the following are therefore entirely due
to the

\narrowtext \begin{figure} 
\epsfxsize=9truecm
\centerline{\epsfbox{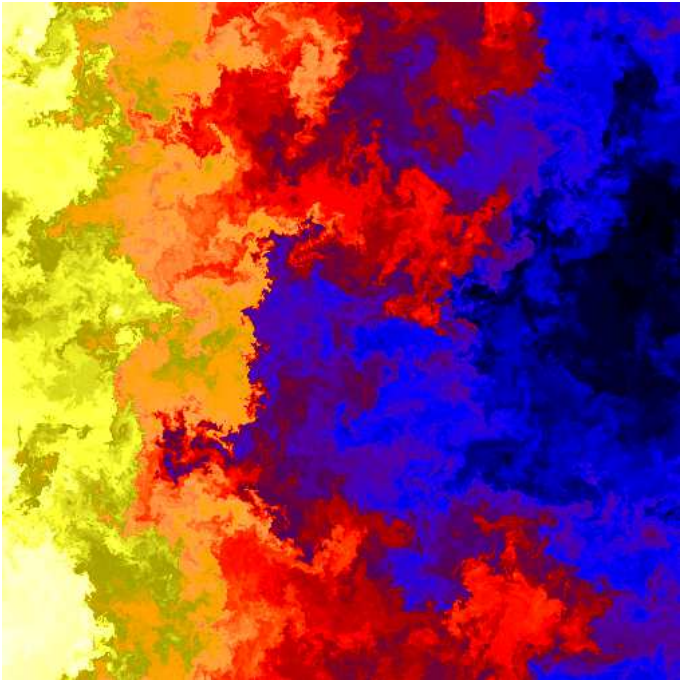}}
\vskip 0.3cm
\caption{A snapshot of the scalar field obtained by numerical
integration of (1) with advection by a two-dimensional turbulent flow
generated by an inverse energy cascade.  The scalar turbulence is
maintained by a fixed temperature gradient $\langle\theta\rangle= {\bm
g}\cdot{\bm r}$, with ${\bm g}$ oriented from right to left.}
\end{figure}

\noindent advection-diffusion equation (\ref{passive}) and not mere
footprints of the velocity field.  Details on the integration
procedure used for the numerical simulations of (\ref{passive}) with a
mean gradient ${\bm g}$ can be found in Ref.~\cite{CLMV00}.  The
single-time scalar statistics at the stationary state is defined by
the $n$-point correlations $C_n(\underline{\bm r},t)=
\langle\theta({\bm r}_1,t)\cdots\theta({\bm r}_n,t)\rangle$, where
$\underline{\bm r}$ denotes the set ${\bm r}_1,\ldots ,{\bm r}_n$.
For spatially homogeneous situations, $C_n$ is invariant under
translations and it depends only on $2n-2$ degrees of freedom,
associated to the separation vectors among the $n$ points. In the
inertial range, the velocity scale invariance is expected to reflect
in scalar correlations $C_n(\underline{\bm r})$ behaving as power laws
with respect to global size variables, such as e.g. the gyration
radius of the set $\underline{\bm r}$.

Let us consider for simplicity the third order case $n=3$.  All the
following arguments are easily generalized to higher order
correlations. The correlation function $C_3$ depends on the size, the
orientation and the shape of the triangle defined by the three points
${\bm r}_1$, ${\bm r}_2$ and ${\bm r}_3$. The global size variable can
be defined as $R^2=(r_{12}^2+r_{23}^2+r_{31}^2)/3$, where $r_{ij}$ is
the distance between the $i$-th and the $j$-th particle.  As shown in
Fig.~2, in the inertial range of scales $C_3$ depends on $R$ as a
power law with the exponent $\zeta_3=1.25$. The hallmark of
intermittency is in the fact that $\zeta_3$ is smaller than the
dimensional prediction $5/3$ (see Ref.~\cite{SS00}). As for the shape
and the orientation of the triangle, we shall use the same Euler
parametrization as in Refs.~\cite{SS98,Pumir}.  Defining ${\bm
\rho}_1=\left({\bm r}_1-{\bm r}_2\right)/\sqrt{2}$ and

\narrowtext \begin{figure}[!hb] 
\epsfxsize=9truecm 
\centerline{\epsfbox{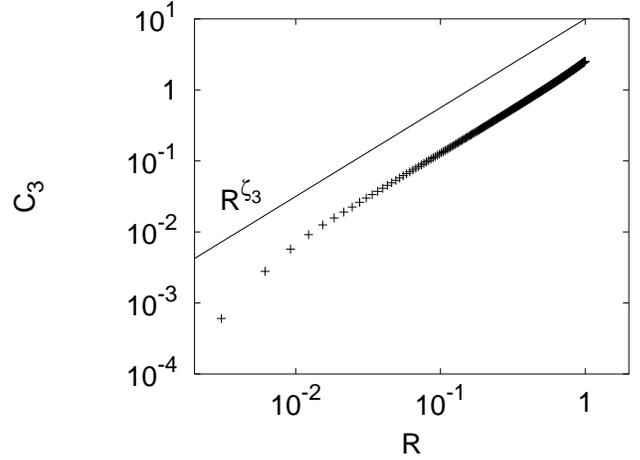}}
\vskip 0.3cm
\caption{The dependence of the third order correlation function $C_3$ 
with respect to the size of the triangle $R$. 
The straight line is the power law behavior $R^{1.25}$.}
\end{figure}
\noindent ${\bm
\rho}_2=\left({\bm r}_1+{\bm r}_2-2{\bm r}_3\right)/\sqrt{6}$, the
shape of the triangle is controlled by the two variables
\begin{equation}
\chi=1/2 \tan^{-1} \left[{2 {\bm \rho}_1
\cdot {\bm \rho}_2 \over (\rho_1^2 - \rho_2^2)}\right]\,;
\quad w=2 {|{\bm
\rho}_1 \times {\bm \rho}_2|\over R}.
\end{equation}
Some of the shapes associated to different values of $\chi$ and $w$
are shown in Fig.~2. The global orientation of the triangle with
respect to the mean gradient direction ${\bm g}$ is defined by the
angle $\varphi$.  It is convenient to decompose $C_3$ on the
orthogonal basis made of $\cos\left(\ell \varphi\right)$ and
$\sin\left(\ell \varphi\right)$.  Reversing the coordinates with
respect to an axis parallel or orthogonal to ${\bm g}$ statistically
leaves the $\theta$ field invariant or inverts its sign,
respectively. In the projection of $C_3$, the angular momentum $\ell$
should therefore be odd and sine functions are absent.  Furthermore,
the dominant contribution at the small scales is the one having the
lowest angular momentum (see Ref.~\cite{Itamar} for the case of
Navier-Stokes turbulence).  The correlation function $C_3$ takes then
the form
\begin{equation}
\label{tre}
C_3(\underline{\bm r})= R^{\zeta_3}\,f(\chi,w)\,\cos\varphi + \ldots\;,
\end{equation}
where the dots stand for subdominant higher-order harmonics of the
form $\cos(2\ell +1)\varphi$.  The invariance under arbitrary
permutations of the three vertices of the triangle allows to reduce
the phase space to $-\pi/6<\chi<\pi/6$, $0<w<1$ and the function $f$
in (\ref{tre}) is antiperiodic in $\chi$ with period $\pi/3$
\cite{SS98,Pumir}.  The measured dependence of $f$ on the shape
coordinates $\chi$ and $w$ is shown in Fig.~3.  The maximum of $f$ is
realized at $\chi=0$, $w=0$, where two of the three particles are
stuck together. For equilateral triangles ($w=1$) or for the
``dumbbell'' configuration $\chi=\pi/6$, $w=0$, the symmetries enforce
$f=0$.

Let us establish the connection with the Lagrangian dynamics. That is
done by replacing the Eulerian variables $\underline{\bm r}$ in the
argument of $C_3$ by their Lagrangian evolutions $\underline{\bm
\rho}(t)$, with $\underline{\bm \rho}(0)=\underline{\bm r}$. The
resulting object is a

\narrowtext \begin{figure}[!hb] 
\epsfxsize=9truecm 
\centerline{\epsfbox{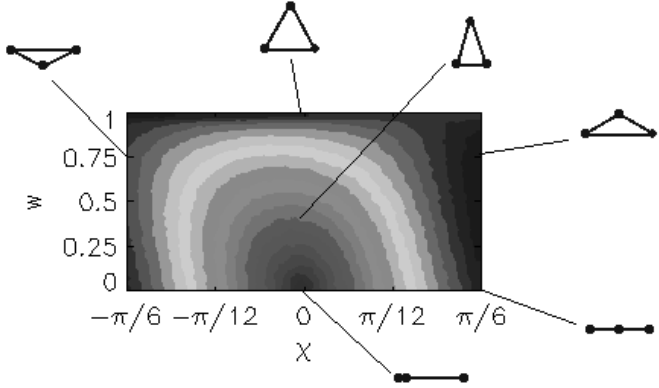}}
\vskip 0.3cm
\caption{Contour lines in the $\chi\,-\,w$ plane of the third order
shape function $f$ appearing in (5).}
\end{figure}
\noindent stochastic function whose average with respect
to the Lagrangian trajectory statistics is denoted by
$\langle\bullet\rangle_{\cal L}$.  More generally, for a generic function
$\phi(\underline{\bm r})$ of the $n$ points ${\bm r}_i$ we can define
its Lagrangian average as
\begin{equation}
\label{evolute}
\langle \phi(t)\rangle_{\cal L}=\int \phi(\underline{\bm \rho})
P_n\left(t\,,\underline{\bm \rho}\big| 0\,,\underline{\bm
r}\right)\,d\underline{\bm \rho}.
\end{equation}
Here, the $n$-particle propagator $P_n\left(t\,,\underline{\bm
\rho}\big|0\,,\underline{\bm r}\right)$ denotes the probability
that, being in $\underline{\bm r}$ at time $0$, the $n$ particles are
in $\underline{\bm \rho}$ at time $t$.  In a turbulent Kolmogorov
flow, distances typically grow with time as $|t|^{3/2}$, whose special
instance is the celebrated Richardson law $\langle
r_{12}^2\rangle_{\cal L}(t)\propto |t|^3$ for the distance $r_{12}$
between two particles. For generic functions $\phi$, homogeneous of
positive degree $\sigma$ that is  $\phi(\lambda\underline{\bm r})=
\lambda^{\sigma}\phi(\underline{\bm r})$, the Lagrangian average is
therefore expected to grow as $|t|^{3\sigma /2}$.

Intermittency is dynamically originated by blatant violations of
expectations {\it \`a la} Richardson. The anomalous part of
correlation functions has in fact a constant Lagrangian average, as
clearly demonstrated in Fig.~4 for the third order case.  Those
statistically preserved functions are the statistical integrals of
motion responsible for the breaking of scale invariance associated to
intermittency.  The other important point is that their constancy is
tightly related to the geometry of the figures identified by the
Lagrangian particles. Fig.~4 indicates indeed that the size factor
$R^{\zeta_3}$ grows as $|t|^{3\zeta_3 /2}$. The Lagrangian average of
$C_3$ remaining constant, the shape part $f(\chi,w)\,\cos\varphi$ in
(\ref{tre}) must compensate for the growth of the figure size. As
shown in Fig.~3, the function $f$ decreases going from degenerate
triangles with two of the vertices close to each other to triangles
with aspect ratios of order unity.  The geometrical meaning of
anomalous scaling laws is then quite clear\,: the smaller is
$\zeta_3$, the slower is the compensation needed from the shape factor
and the longer degenerate triangle configurations persist.
In other words, ``stronger intermittency = particles staying
longer close to each other''.

\narrowtext \begin{figure}[h] \epsfxsize=9truecm
\centerline{\epsfbox{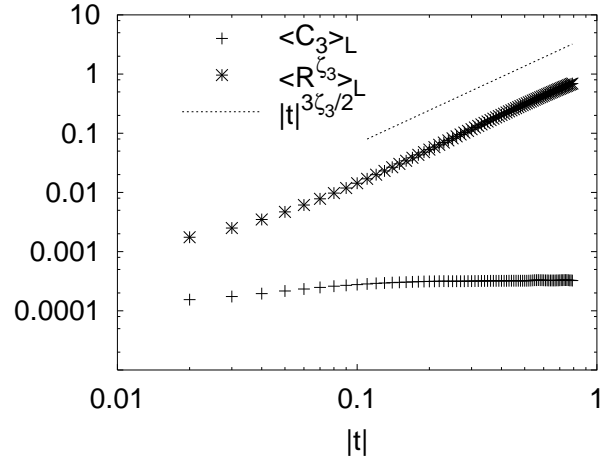}}
\vskip 0.3cm
\caption{The Lagrangian average of the correlation function $\langle
C_3 \rangle_{\cal L}$. For comparison, it is also shown the evolution
of the average $\langle R^{\zeta_3}\rangle_L \sim |t|^{{3 \over
2}\zeta_3}$, that obeys the dimensional scaling law. }
\end{figure}

Systematic support to the previous physical ideas can be provided in
the special case where the advecting velocity in (\ref{passive}) has a
short correlation time, the so-called Kraichnan model \cite{RHK94}.
The assumption is of course far from realistic, but it leads to the
peculiar property that the $n$-particle propagators $P_n$ in
(\ref{evolute}) obey closed Fokker-Planck equations \cite{RHK68} (see
also Ref.~\cite{SS00}). The statistically preserved functions are now
identified as zero modes of the $n$-particle Fokker-Planck operator
\cite{CFKL95,GK95,SS95}. Their anomalous scaling behavior could be
calculated in some perturbative limits \cite{CFKL95,GK95,SS95} or
measured numerically \cite{FMV,GPZ}.  Zero modes enter the $P_n$'s via
the asymptotic expansion \cite{BGK} (see also Ref.~\cite{GZ})\,:
\begin{equation}
\label{closepart}
P_n\!\!\left(t,\underline{\bm \rho}\Big| 0,\lambda\,\, \underline{\bm r}
\right)\!\!\!  =\!\!  \sum_{i,q}\,\, \lambda^{\sigma_{i,q}}
\phi_{i,q}(\underline{\bm r})\, \psi_{i,q} (t, \underline{\bm \rho}),
\end{equation}
valid for small $\lambda$'s. Zero modes are the $q=0$ terms in
(\ref{closepart}) and they are ordered according to their scaling
dimension by the index $i$. Higher $q$'s identify the so-called slow
modes, whose Lagrangian average is growing as an integer power of
time, although with an exponent smaller than the dimensional one
$3\sigma_{i,q}/2$.

A simple example of slow mode for our inverse cascade velocity field
is provided for $n=2$. The Lagrangian average of $\left({\bm
g}\cdot{\bm r}_{12}\right)$ is preserved as its time derivative is
proportional to $\langle \left({\bm v}_1- {\bm
v}_2\right)\rangle=0$. The first slow mode associated to it is given
by the function $\left({\bm g}\cdot{\bm
r}_{12}\right)\,{r_{12}^{2/3}}$. Fig.~5 shows indeed that its
Lagrangian average grows as $|t|$, much slower at large
times than the dimensional law $|t|^{5/2}$.

What is the degree of generality of the Lagrangian preservation
mechanism for intermittency and the expansion (\ref{closepart})\,?
The crucial property of the Kraichnan velocity 
field is its short correlation time (this ensures that the

\narrowtext \begin{figure}[h] \epsfxsize=9truecm
\centerline{\epsfbox{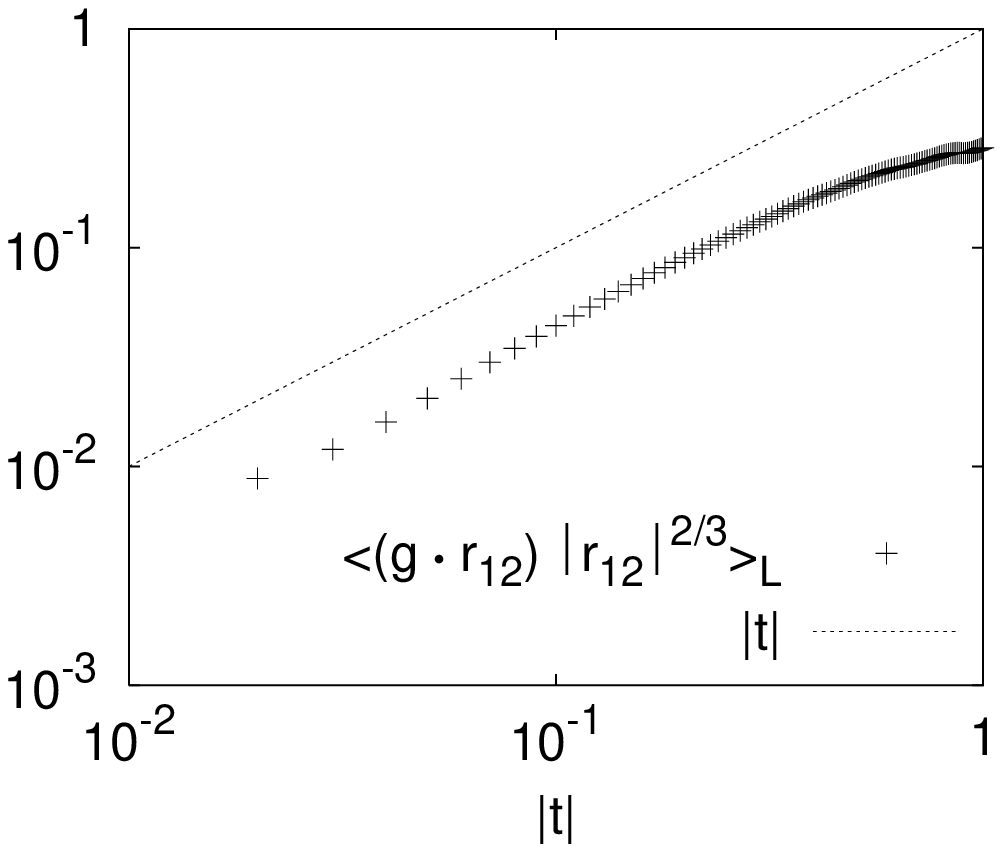}}
\vskip 0.3cm
\caption{The Lagrangian average of the anisotropic slow
mode $\left({\bm g}\cdot{\bm r}_{12}\right)\,{r_{12}^{2/3}}$
{\it vs} time.}
\end{figure}

\noindent Lagrangian trajectories are Markovian). For the inverse
energy cascade flow considered here this property is lost as the
correlation time is finite. The numerical results presented here give
therefore a strong indication that the basic mechanisms for scalar
intermittency are quite robust and generic. Changing the statistics of
the flow affects only quantitative details, such as the numerical
value of the exponents or the precise shape of the Lagrangian
preserved functions.  The expansion (\ref{closepart}) is analogous to
those encountered in many-body statistical problems \cite{foot} and
controls the small-scale behavior of scalar correlations. Indeed, it
follows from (\ref{passive}) that the scalar is preserved along the
Lagrangian trajectories. Scalar correlations can be then expressed as
\begin{equation}
\label{corre}
C_n(\underline{\bm r},t)= \int P_n\left(-t,\underline{\bm \rho}\big|
0,\underline{\bm r}\right)\left({\bm g}\cdot{\bm \rho}_1\right)\ldots
\left({\bm g}\cdot{\bm \rho}_n\right) d\underline{\bm \rho} .
\end{equation}
Inserting (\ref{closepart}) into this expression it is evident that
the behavior at small scales is a superposition of the functions
$\phi_{i,q}$. Two general remarks on scalar turbulence follow. First,
changing the initial condition ${\bm g}\cdot{\bm r}$ (or the injection
mechanism in the forced case) modifies the constants but not the
scaling exponents in the correlation functions. This naturally
explains the universality properties of scalar turbulence observed in
Ref.~\cite{CLMV00}.  Second, all fewer-particle modes $\phi_{i,q}$,
i.e. those depending on $m<n$ variables, appear in the expression for
$C_n$. Their Lagrangian averages with respect to the evolution of the
$m$ particles or the whole set of $n$ particles are indeed trivially
coinciding. Note however that structure functions satisfy the trivial
identity $S_n(r)= \int_0^r\ldots \int_0^r
\partial_{r_1}\ldots\partial_{r_n}C_n(\underline{r})\,d\underline{r}$.
All $m$-particle modes will therefore drop out from the $n$-th order
structure function. This clearly illustrates a point previously
mentioned\,: the apparent simplification of the single distance $r$
left in structure functions is purely illusory as their anomalous
scaling laws are still dynamically associated to $n$-particle
geometries.

In conclusion, we have identified the integrals of motion responsible
for the failure of dimensional analysis and intermittency in scalar
turbulence. The anomalous parts of correlation functions are
preserved under the Lagrangian dynamics.  That preservation is a
geometrical effect associated to figures identified by scalar
particles persisting in degenerate shapes.  Particles tend indeed to
remain close to each other for unexpectedly long times, pointing to
applications for the reaction rate enhancement in the transport of
chemically reactive species.

\medskip {\bf Acknowledgments} We are grateful to U.~Frisch and
K.~Gaw\c{e}dzki for illuminating discussions. A.~Celani was
supported by a ``H.~Poincar\'e'' CNRS postdoctoral fellowship. The
work was partially supported by the EU contract HPRN-CT-2000-00162 
``Nonideal Turbulence'' and the NSF under Grant No. PHY94-07194.

\end{document}